\documentclass[prd,twocolumn,superscriptaddress,amsfonts,amssymb,amsmath,showpacs]{revtex4-2}
\usepackage{amssymb}
\usepackage{bm}
\usepackage{amsfonts}
\usepackage{latexsym}
\usepackage[latin1]{inputenc}
\usepackage{graphicx}
\usepackage{amsmath}
\usepackage{palatino}
\usepackage{mathpazo}
\usepackage{textcomp}
\usepackage{float}
\usepackage{booktabs}
\usepackage{caption}
\usepackage{subcaption}
\usepackage{dcolumn}
\usepackage{ragged2e}
\usepackage{hyperref}
\hypersetup{colorlinks,citecolor=blue}
 
\usepackage[many]{tcolorbox} 
\newtcolorbox{boxB}{
    fontupper = \bf, 
    boxrule = 1.5pt,
     width = 18cm,
    colframe = black,
    colback = white,
    rounded corners,
    arc = 5pt   
    }

\usepackage{amsmath}
\usepackage{xcolor}
\usepackage{orcidlink}
\usepackage{epsfig}
\usepackage{commath}
\usepackage{tabularx}
\usepackage{multirow}
\usepackage{mathtools}
\newcommand{\be}{\begin{equation}}
	\newcommand{\ee}{\end{equation}}
\newcommand{\bea}{\begin{eqnarray}}
	\newcommand{\eea}{\end{eqnarray}}
\newcommand{\beas}{\begin{eqnarray*}}
	\newcommand{\eeas}{\end{eqnarray*}}

\allowdisplaybreaks[1]
\renewcommand{\arraystretch}{1.1}
\begin{document}
	\color{black} 
	\title{Stable $f(Q)$ gravity model through non-trivial connection}
	
	\author{S.A. Narawade \orcidlink{0000-0002-8739-7412}}
	\email{shubhamn2616@gmail.com}
	\affiliation{Department of Mathematics, Birla Institute of Technology and
		Science-Pilani,\\ Hyderabad Campus, Hyderabad-500078, India.}
        \author{Santosh V. Lohakare \orcidlink{0000-0001-5934-3428}}
        \email{lohakaresv@gmail.com}
        \affiliation{Department of Mathematics, Birla Institute of Technology and
		Science-Pilani,\\ Hyderabad Campus, Hyderabad-500078, India.}
	\author{B. Mishra \orcidlink{0000-0001-5527-3565}}
	\email{bivu@hyderabad.bits-pilani.ac.in}
	\affiliation{Department of Mathematics, Birla Institute of Technology and
		Science-Pilani,\\ Hyderabad Campus, Hyderabad-500078, India.}

\begin{abstract}
    \textbf{Abstract:} This study effectively reconstructs a cosmological model utilizing covariant $f(Q)$ gravity within Connection-III and FLRW spacetime. The dynamic behavior of the reconstructed model is thoroughly analyzed using the Hubble parameter $H(z)$ and various observational datasets. Our robust findings demonstrate that the model displays quintessence behavior at the present epoch and converges to the $\Lambda$CDM model at late time. It is confirmed through comprehensive evaluations against energy conditions that the Null Energy Condition remains positive throughout cosmic evolution, and the Dominant Energy Condition is consistently satisfied. The Strong Energy Condition is initially fulfilled in the early Universe but violated in the late epoch. Moreover, scalar perturbations extensively assess stability, affirming the strength of the model with respect to the Hubble parameter. This research offers compelling insights into cosmic acceleration, suggesting that $f(Q)$ gravity can effectively displace the $\Lambda$CDM model and provides a convincing alternative explanation for the current accelerating expansion of the Universe without relying on the cosmological constant.

\end{abstract}		

\maketitle

\section{Introduction}\label{Sec:I}
    
    A significant challenge in modern cosmology is to explain the origin and existence of the dark components of the Universe, which account for the observed acceleration of the Universe \cite{Riess1998, Perlmutter1999}. This finding was later confirmed by additional data from the cosmic microwave background (CMB) \cite{Page2003, Spergel2007, Komatsu2009, Komatsu2011} and large-scale structure (LSS) \cite{Tegmark2004, Seljak2005}. This discovery has become a pivotal development in physics and astronomy. The cosmological constant, characterized by a negative equation of state (EoS), offers the simplest explanation for dark energy, which is responsible for this accelerated expansion. Recently, modified gravity theories such as $f(R)$ and $f(T)$ \cite{Sotiriou_2010_82, Ferraro_2007_75, Nojiri2011, DeFelice2010, Linder2010, D'Agostino2020} have attracted considerable interest as alternative approaches to understanding dark energy \cite{Caldwell1998, Boshkayev2019}. As an alternative, there is also the nonmetricity approach, in which a nonmetricity $Q$ mediates gravitational interaction, free of curvature and torsion, known as symmetric teleparallel gravity \cite{Nester1999}. The symmetric teleparallel gravity has developed into coincident general relativity (GR) or $f(Q)$ gravity by Jimenez et al. \cite{Jimenez2018}. Since modern cosmology has emphasized extended gravity theories, alternative geometries are also being studied. With the belief that gravitational theories would have different properties corresponding to Riemannian geometry, we are attempting to gain insight into cosmic acceleration at late time. The concordance $\Lambda$CDM model can be effectively replaced by $f(Q)$ gravity.

    The non-minimally coupled $f(Q)$ cosmology is proposed to explain the current accelerating expansion of the Universe in addition to the examples of minimally coupled matter without using the cosmological constant \cite{Harko2018}. In symmetric teleparallel gravity, the choice of connection is crucial as it determines the nonmetricity scalar $Q$, this impacts the equations of motion and influences the cosmological dynamics. Various connections provide unique insights into the relationship between geometry and cosmic evolution. For example, Connection I corresponds to coincident gauge choices, ensures a more straightforward dynamic formulation. In contrast, Connection II and Connection III introduce more complex dynamical characteristics, allowing for a deeper investigation into cosmological phenomena. Recently, $f(Q)$ gravity and its cosmological implications have been the subject of several important publications, see \cite{Lazkoz2019, Lu2019, Jimenez2020, Barros2020, Frusciante2021, Anagnostopoulos2021, Khyllep2021, Lin2021, Narawade2022, D'Ambrosio2022, Narawade2023a, Capozziello_2022_82, Capozziello_2023_83, Jensko_2024_2407.17568}. It is worth noting that all of these studies were conducted with coincident gauges and line elements in Cartesian coordinates. Due to this specific choice, the covariant derivative is reduced to a partial derivative, which simplifies the calculations. On the other hand, the pressure and energy equations are identical to the $f(T)$ theory. In cosmological contexts, whether in flat \cite{Subramaniam2023, Shabani2023, Paliathanasis2023} or curved \cite{Dimakis2022, Heisenberg2023, Shabani2023a, Subramaniam2023a} Universe, the $f(Q)$ theory, which does not rely on coincident gauges, is garnering significant interest. 

    The reconstruction of cosmology is an essential tool in modified gravity for imitating realistic cosmological scenarios. Researchers have conducted reconstruction schemes in $f(R)$ gravity and its modifications under various scenarios \cite{Elizalde_2004_70, Cognola_2005_04, Nojiri_2006_74_086005, Capozziello_2005_71, Capozziello_2006_73, Nojiri_2010_42, Li_2011_83} to discover a realistic cosmology that can account for the transition from the matter-dominated epoch to the dark energy phase. One approach involves considering the known cosmic evolution and utilizing the field equations to derive a specific form of Lagrangian that replicates the given background evolution. Nojiri et al. \cite{Nojiri_2009_681, Bahamonde_2015_01_186} undertook a reconstruction scheme to identify realistic models in the $f(R)$ theory, which was subsequently extended to $f(R, \mathcal{G})$ modified Gauss-Bonnet theories \cite{Elizalde_2010_27} (where $\mathcal{G}$ is the Gauss-Bonnet term). Capozziello and D'Agostino investigated the model-independent reconstruction of $f(Q)$ non-metric gravity \cite{Capozziello_2022_832}. Additionally, the cosmic evolution based on the power law solution of the scale factor has been examined in modified theories \cite{Goheer_2009_79}. 

    The research investigates whether $f(Q)$ gravity could be a substitute for the standard cosmological model. This is done by reconstructing models, leading to a promising $f(Q)$ model that effectively represents the current cosmic epochs and monitors how energy components change over time. To evaluate the cosmological implications of the model, we conducted an extensive analysis of its background dynamics using Markov Chain Monte Carlo (MCMC) methods. We applied this analysis to late-time observational data, including Supernovae Ia, observational Hubble data and BAO data. Simultaneously, the study investigated the stability of the cosmological background within the $f(Q)$ gravity framework under the assumptions of homogeneity and isotropy. Using a reconstruction approach, we constructed the $f(Q)$ model and tested its stability using a scalar perturbation analysis. Ongoing research aims to derive analytical solutions for the evolution of perturbation variables and ultimately determine the evolution function for the reconstructed model. To explore the potential impact of nonmetricity on the evolution of the universe, we constructed an $f(Q)$ model that incorporates nontrivial connections. By comparing this model to the $\Lambda$CDM paradigm, the goal is to assess the role of nonmetricity in cosmic dynamics.

    The paper is organized as follows: In Sec. \ref{Sec:II}, we provide a mathematical formulation of symmetric teleparallelism with a focus on the $f(Q)$ theory, and we describe all of the affine connections in the spatially flat FLRW spacetime. By applying specific conditions along with connection-III, we reconstruct the $f(Q)$ model in Sec. \ref{Sec:III}. In Sec. \ref{Sec:IV}, we analyse the dynamical behavior and the stability through scalar perturbation for the reconstructed model. We conclude with a general discussion of our results in Sec. \ref{Sec:V}.

\section{$f(Q)$ Gravity: Covariant Formulation}\label{Sec:II}
The torsion tensor $\mathcal{T}^{\lambda}_{~~\mu\nu}$, as well as the nonmetricity tensor $Q_{\lambda\mu\nu}$, are respectively presented for a spacetime equipped with the metric tensor $g_{\mu\nu}$ and the affine connection $\Gamma^{\lambda}_{~~\mu\nu}$.
\begin{eqnarray}\label{eq1}
    \mathcal{T}^{\lambda}_{~~\mu\nu} &=& \Gamma^{\lambda}_{~~\mu\nu}-\Gamma^{\lambda}_{~~\nu\mu},\nonumber\\
    Q_{\lambda\mu\nu} &=& \nabla_{\lambda}g_{\mu\nu} = \partial_{\lambda}g_{\mu\nu}-\Gamma^{\alpha}_{~~\lambda\mu}g_{\alpha\nu}-\Gamma^{\alpha}_{~~\lambda\nu}g_{\alpha\mu}.
\end{eqnarray}
An affine connection $\Gamma^{\lambda}_{~~\mu\nu}$ is related to a Levi-Civita connection $\{^{\lambda}_{~\mu \nu}\}$ by,
\begin{equation}\label{eq2}
    \Gamma^{\lambda}_{~~\mu\nu} = \{^{\lambda}_{~\mu \nu}\} + \mathcal{S}^{\lambda}_{~~\mu\nu}~,
\end{equation}
where, $\{^{\lambda}_{~\mu\nu}\} = \frac{1}{2} g^{\lambda \alpha}\left({\partial}_{\mu} g_{\alpha \nu}+{\partial}_{\nu} g_{\alpha \mu}-{\partial}_{\alpha} g_{\mu \nu}\right)$ and the distortion tensor is $\mathcal{S}_{\lambda\mu\nu} = -\frac{1}{2}\left(\mathcal{T}_{\mu\nu\lambda}+\mathcal{T}_{\nu\mu\lambda}-\mathcal{T}_{\lambda\mu\nu} \right) -\frac{1}{2}\left(Q_{\mu\nu\lambda}+Q_{\nu\mu\lambda}-Q_{\lambda\mu\nu} \right)$. The nonmetricity scalar is defined as $Q = Q_{\lambda\mu\nu}P^{\lambda\mu\nu}$, its two traces as $Q_{\alpha} = Q_{\alpha~~\mu}^{~~\mu}$ and $\tilde{Q}_{\alpha}=Q^{\mu}_{~\alpha\mu}$. The $P^{\lambda}_{~~\mu\nu}$ is called as nonmetricity conjugate and given as 
\begin{multline}\label{eq3}
   P^{\lambda}_{~~\mu\nu} = -\frac{1}{4}Q^{\lambda}_{~\mu \nu} + \frac{1}{4}\left(Q^{~\lambda}_{\mu~\nu} + Q^{~\lambda}_{\nu~~\mu}\right) + \frac{1}{4}Q^{\lambda}g_{\mu \nu} \\
   - \frac{1}{8}\left(2 \tilde{Q}^{\lambda}g_{\mu \nu} + {\delta^{\lambda}_{\mu}Q_{\nu} + \delta^{\lambda}_{\nu}Q_{\mu}} \right).
\end{multline}
A different definition of $Q = -Q_{\lambda\mu\nu}P^{\lambda\mu\nu}$ has been proposed in the literature, which changes the sign of the nonmetricity scalar $Q$. This is important to consider while comparing different $f(Q)$ results. Symmetric teleparallel equivalent of GR (STEGR) can be produced if this nonmetricity scalar $Q$ replaces the Ricci scalar $R$ in the Einstein-Hilbert action. It is noteworthy that STEGR faces the same dark problem as in the conventional GR. Modified gravity theories of the form $f(Q)$ were developed to address this issue, in a similar way to the extensions seen in modified gravity theories represented by $f(R)$. By varying the action term \cite{Jimenez2018}
\begin{equation}\label{eq4}
    S = \int \frac{1}{2\kappa}f(Q)\sqrt{-g}~d^{4}x + \int \mathcal{L}_{m}\sqrt{-g}~d^{4}x,
\end{equation}
with respect to the metric tensor, we can obtain the field equation
\begin{multline}\label{eq5}
    \frac{2}{\sqrt{-g}}\nabla_{\lambda}\left(\sqrt{-g}f_{Q}P^{\lambda}_{~~\mu\nu}\right) - \frac{1}{2}g_{\mu \nu}f \\
    + f_{Q}(P_{\mu\lambda\alpha}Q^{~~\lambda \alpha}_{\nu} - 2Q_{\lambda \alpha \mu}P^{\lambda \alpha}_{~~~\nu}) = \kappa T_{\mu \nu}~.
\end{multline}
Using this field equation, the covariant formulation has been developed and used effectively in studying geodesic deviations and cosmological phenomena \cite{Beh2022, Zhao2022},
\begin{equation}\label{eq6}
    f_{Q}\overcirc{G}_{\mu\nu}+\frac{1}{2}g_{\mu\nu}(Qf_{Q}-f)+2f_{QQ}P^{\lambda}_{~~\mu\nu}\overcirc{\nabla}_{\lambda}Q = \kappa T_{\mu \nu}~,
\end{equation}
where,$\overcirc{\nabla}$ is the covariant derivative associated with the Levi-Civita connection, $f_{Q}$ is derivative of $f$ with respect to $Q$ and $\overcirc{G}_{\mu\nu} = R_{\mu\nu}-\frac{1}{2}g_{\mu\nu}R$, with $R_{\mu\nu}$ and $R$ are the Ricci tensor and scalar respectively which are constructed by the Levi-Civita connection. For a linear form of $f(Q)$ function, the above equation reduces to GR. Variation of Eq. \eqref{eq4} with respect to the connection, we can derive the equation of motion for the nonmetricity scalar as,
\begin{equation}\label{eq7}
    \nabla_{\mu}\nabla_{\nu}\left(\sqrt{-g}f_{Q}P^{\mu\nu}_{~~~\lambda}\right)=0.
\end{equation}
\par The affine connection discussed here is just a special case of one of the three classes discussed in the next paragraph. So we must look beyond this particular gauge choice. We consider the spatially homogeneous, isotropic, and flat FLRW space-time for the study of background geometry,
\begin{equation}\label{eq8}
    ds^2=-dt^2+a(t)^2\left[dr^{2}+r^{2}\left(d\theta^{2}+sin^{2}\theta ~d\phi^{2}\right)\right],
\end{equation}
where $a(t)$ is scale factor and the Hubble function defined as $H= \frac{\dot{a}}{a}$ with $\dot{a}$ is derivative of $a$ with respect to time $t$. These connections can be classified into three types, based on $K_{1}$, $K_{2}$ and $K_{3}$ given by \cite{Dimakis2022b, Zhao2022, Shi2023}.

\begin{eqnarray}\label{eq9}
 \Gamma^t_{~tt} &=& K_{1}, \quad \Gamma^t_{~rr} = K_{2}, \quad \Gamma^t_{~\theta\theta} = K_{2}r^2, \nonumber\\
 \Gamma^r_{~tr} &=& \Gamma^r_{~rt} = \Gamma^{\theta}_{~t\theta} = \Gamma^{\theta}_{~\theta t} = \Gamma^\phi_{~t\phi} = \Gamma^\phi_{~\phi t} = K_{3}, \nonumber\\
 \Gamma^\theta_{~r\theta} &=& \Gamma^\theta_{~\theta r} = \Gamma^\phi_{~r\phi} = \Gamma^\phi_{~\phi r} = \frac{1}{r}, \quad \Gamma^r_{~\theta\theta} = -r, \nonumber\\
 \Gamma^r_{~\phi\phi} &=&  -r\sin^2\theta, \quad \quad \Gamma^t_{~\phi\phi} = K_{2}r^2\sin^2\theta, \nonumber\\
 \Gamma^\phi_{~\theta\phi} &=& \Gamma^\phi_{~\phi\theta} = \cot\theta, \quad \Gamma^{\theta}_{~\phi\phi} = -\cos\theta\sin\theta~.
\end{eqnarray}
To derive the nonmetricity scalar $Q$, we use the definitions provided in Eqs. \eqref{eq1} and \eqref{eq9}. The expression for 
$Q$ is given by:
\begin{multline}\label{eq10}
    Q = -6H^{2}+9HK_{3}+3K_{3}\left(K_{1}-K_{3}\right)\\
    +\frac{3K_{2}H}{a^{2}}-\frac{3K_{2}\left(K_{1}+K_{3}\right)}{a^{2}},
\end{multline}
We consider three different connections, each defined as follows:
\begin{itemize}
    \item Connection I: $K_{1} = \gamma(t), \quad K_{2} = K_{3} = 0$;    
    \item Connection II: $K_{1} = \gamma(t) + \frac{\dot{\gamma}(t)}{\gamma(t)}, \quad K_{2} = 0, \quad K_{3} = \gamma(t)$;  
    \item Connection III: $K_{1} = -\frac{\dot{\gamma}(t)}{\gamma(t)}, \quad K_{2} = \gamma(t), \quad K_{3} = 0$~,
\end{itemize}
where $\gamma$ is the non-vanishing function of $t$. For each connection, the resulting field equations are different. The nonmetricity scalar $Q$ and Friedmann equations for each connection are:\\
\textbf{For Connection I:}
\begin{eqnarray}
    && Q = -6H^{2}~, \label{eq11}\\
    && 3H^{2}f_{Q}+\frac{1}{2}(f-Qf_{Q}) = \rho~,\label{eq12}\\
    && -2\frac{d(f_{Q}H)}{dt}-3H^{2}f_{Q}-\frac{1}{2}(f-Qf_{Q}) = p~,\label{eq13}
\end{eqnarray}
where $\rho$ denotes the energy density, including baryonic matter, cold dark matter, and radiation, and $p$ represents the pressure.\\
\textbf{For Connection II:}
\begin{eqnarray}
    && Q = -6H^{2}+9\gamma H+3\dot{\gamma}~, \label{eq14}\\
    && 3H^{2}f_{Q}+\frac{1}{2}(f-Qf_{Q})+\frac{3\gamma}{2}\dot{Q}f_{QQ} = \rho~,\label{eq15}\\
    && -2\frac{d(f_{Q}H)}{dt}-3H^{2}f_{Q}-\frac{1}{2}(f-Qf_{Q})+\frac{3\gamma}{2}\dot{Q}f_{QQ} = p~,\nonumber\\
    \label{eq16}
\end{eqnarray}
\textbf{For Connection III:}

\begin{eqnarray}
    && Q = -6H^{2}+\frac{3\gamma H}{a^{2}}+\frac{3\dot{\gamma}}{a^{2}}, \label{eq17}\\
    && 3H^{2}f_{Q}+\frac{1}{2}(f-Qf_{Q})-\frac{3\gamma}{2a^{2}}\dot{Q}f_{QQ} = \rho~,\label{eq18}\\
    && -2\frac{d(f_{Q}H)}{dt}-3H^{2}f_{Q}-\frac{1}{2}(f-Qf_{Q})+\frac{\gamma}{2a^{2}}\dot{Q}f_{QQ} = p~.\nonumber\\
    \label{eq19}
\end{eqnarray}

In Connection I, the nonmetricity scalar $Q$ and the Friedmann equations \eqref{eq11} - \eqref{eq13} are independent of the function $\gamma$, aligning with the results obtained under the coincident gauge where $\Gamma^{\lambda}_{~~\mu\nu}=0$ in the flat FLRW metric.

\section{Reconstruction of the $f(Q)$ Model: Analytical Formulation}\label{Sec:III}
 Connection I corresponds to coincident gauge choices, it simplifies the dynamical formulation. While Connection II and Connection III share a similar structure, the function $\gamma$ which acts as an additional degree of freedom, behaves differently in each case \cite{Guzman_2024_110_124013}. The $\frac{1}{a^2}$ factor in Connection III plays a crucial role by suppressing the contributions of $\gamma$ and $\dot{\gamma}$ at late times, ensuring a smoother cosmological evolution and reducing the risk of sudden singularities compared to Connection II. This scaling enables a clear distinction between early and late-time dynamics, making the model more stable and better aligned with observational data. Thus we will focus on Connection III for further analysis. By employing $\gamma = \gamma_{1}a^{2}(t)$ \cite{Subramaniam2023}, where $\gamma_{1}$ is arbitrary constant. We derive $Q=-6H^{2}+9\gamma_{1}H$, which leads to $\dot{Q} = -12H\dot{H}+9\gamma_{1}\dot{H}$. Consequently, Eqs. \eqref{eq18} and \eqref{eq19} can be rewritten as follows,
\begin{eqnarray}
    \rho &=& \frac{f}{2}+3H^{2}f_{Q}-\frac{1}{2}Qf_{Q}-\frac{3\gamma_{1}}{2}\dot{Q}f_{QQ}~,\label{eq20}\\
    p &=& -\frac{f}{2}-3H^{2}f_{Q}+\frac{1}{2}Qf_{Q}-2\dot{H}f_{Q}+\frac{(\gamma_{1}-4H)}{2}\dot{Q}f_{QQ}~.\nonumber\\
    \label{eq21}
\end{eqnarray}
From Eqs. \eqref{eq20} and \eqref{eq21}, we derive the expression,
\begin{equation}\label{eq22}
    \rho + p = \left(-\gamma_{1}\dot{Q}f_{QQ}-2\dot{H}f_{Q}-2H\dot{Q}f_{QQ}\right).
\end{equation}
By applying the chain rule to $f_{Q}$ and considering the condition $p+\rho\rightarrow 0$ we can obtain,
\begin{eqnarray*}
    -\frac{d}{dt}\left(\gamma_{1}f_{Q}\right) &=&  \frac{d}{dt}\left(2Hf_{Q}\right)~,\\
    \therefore\frac{df}{dH} &=& \frac{C(4H-3\gamma_{1})}{2H+\gamma_{1}}~.
\end{eqnarray*}
Upon integrating both sides, we can find
\begin{eqnarray*}
    f(H) &=& 4H-5\gamma_{1}\,\text{ln}(2H+\gamma_{1})~.
\end{eqnarray*}
By applying these results up to the $4^{th}$ order and substituting the value of $Q$, we can reconstruct $f(Q)$ as,
\begin{eqnarray}\label{eq23}
     f(Q) = ~\alpha_{1}+\alpha_{2}Q+\alpha_{3}Q^{2}\nonumber+\left(\beta_{1}+\beta_{2}Q\right)\sqrt{81\gamma_{1}^{2}-24Q}~.\nonumber\\
\end{eqnarray}

\subsection{Data Analysis and Parameter Estimation}
In this section, we analyze the behavior of the $f(Q)$ model by utilizing the functional form of the Hubble parameter given by $H(z)^{2} = H_{0}^2\left[A(1+z)^{B}+\sqrt{A^{2}(1+z)^{2B}+C}\right]$ \cite{Sahni2003, Narawade2023c}. We will conduct MCMC analysis using the \textit{emcee} package \cite{Foreman-Mackey2013}, to constrain the free parameters 
$H_{0}$, $A$, $B$ and $C$ based on observational datasets. The one-dimensional distributions will illustrate the posterior distributions of each parameter, while the two-dimensional distributions will show the covariance between parameter pairs. Our analysis will use Hubble data, Pantheon$^{+}$ data and BAO data as the baseline.\\

\noindent\textbullet\textbf{{Hubble Data:}} The Hubble data were determined using $31$ data points that were derived via the cosmic chronometers (CC) technique. Based on \cite{Zhang2014, Jimenez2003, Moresco2016, Simon2005, Moresco2012, Stern2010, Moresco2015, Moresco2022}, we adopted CC data points that are independent of the Cepheid distance scale and from any cosmological model, but are dependent on stellar ages which are modeled using robust stellar population synthesis techniques (for CC systematic analyses, see Refs. \cite{Moresco2016, Moresco2012, Gomez-Valent2018, Lopez-Corredoira2017, Lopez-Corredoira2018, Verde2014}). The Hubble chi-square ($\chi^2_{H}$) can be calculated with,
\begin{equation}\label{eq24}
 \chi^{2}_{H}(P_s)=\sum_{i=1}^{31} \frac{[H_{th}(z_i,P_s)-H_{obs}(z_i)]^2}{\sigma_H^2(z_i)},
 \end{equation}
where $H_{obs}(z_i)$ represents the observed Hubble parameter values, $H_{th}(z_i,P_s)$ represents the Hubble parameter with the model parameters and $\sigma_H^2(z_i)$ is the standard deviation.

\noindent\textbullet{\textbf{Pantheon$^{+}$ Data:}} Additionally, we have in our baseline data set the Pantheon$^+$ sample data set consists of $1701$ light curves of $1550$ distinct Type Ia supernovae, ranging in redshift from $z = 0.00122$ to $2.2613$ \cite{Brout2022}. The model parameters are fitted by comparing the observed and theoretical values of the distance moduli. Henceforth, we will be denoting the Pantheon$^+$ compilation by \textit{SN}. In our MCMC analyses, we will consider \textit{M} as a nuisance parameter since each SNIa's apparent magnitude must be calibrated through an arbitrary fiducial absolute magnitude \textit{M}. This can be achieved by comparing theoretical distance moduli to actual distance moduli, which can be defined as,
\begin{equation}\label{eq25}
\mu(z_{i},\Theta)=5\log_{10}\big[D_{L}(z_{i},\Theta)\big]+M,
\end{equation}
at redshift $z_{i}$ via the corresponding computation of the luminosity distance
\begin{equation}\label{e26}
D_{L}(z_{i},\Theta) = c(1+z_{i})\int_{0}^{z_{i}}\frac{d\tilde{z}}{H(\tilde{z},\Theta)},
\end{equation}
where $c$ is the speed of light and the associated ($\chi^2_{SN}$) is specified by,
\begin{equation}\label{eq27}
\chi^2_{SN} = \big(\Delta\mu(z_i, \Theta)\big)^T C^{-1}\big(\Delta\mu(z_i, \Theta)\big),
\end{equation}
where $\Delta\mu(z_i, \Theta) = \mu(z_i, \Theta) - \mu_{obs}(z_i)$ is the difference between the predicted and observed distance moduli; $C$ is the corresponding covariance matrix that accounts for statistical and systematic uncertainties.

\begin{figure*}[!htb]
    \centering
    \begin{subfigure}[b]{0.65\textwidth}
        \centering
        \includegraphics[width=105mm]{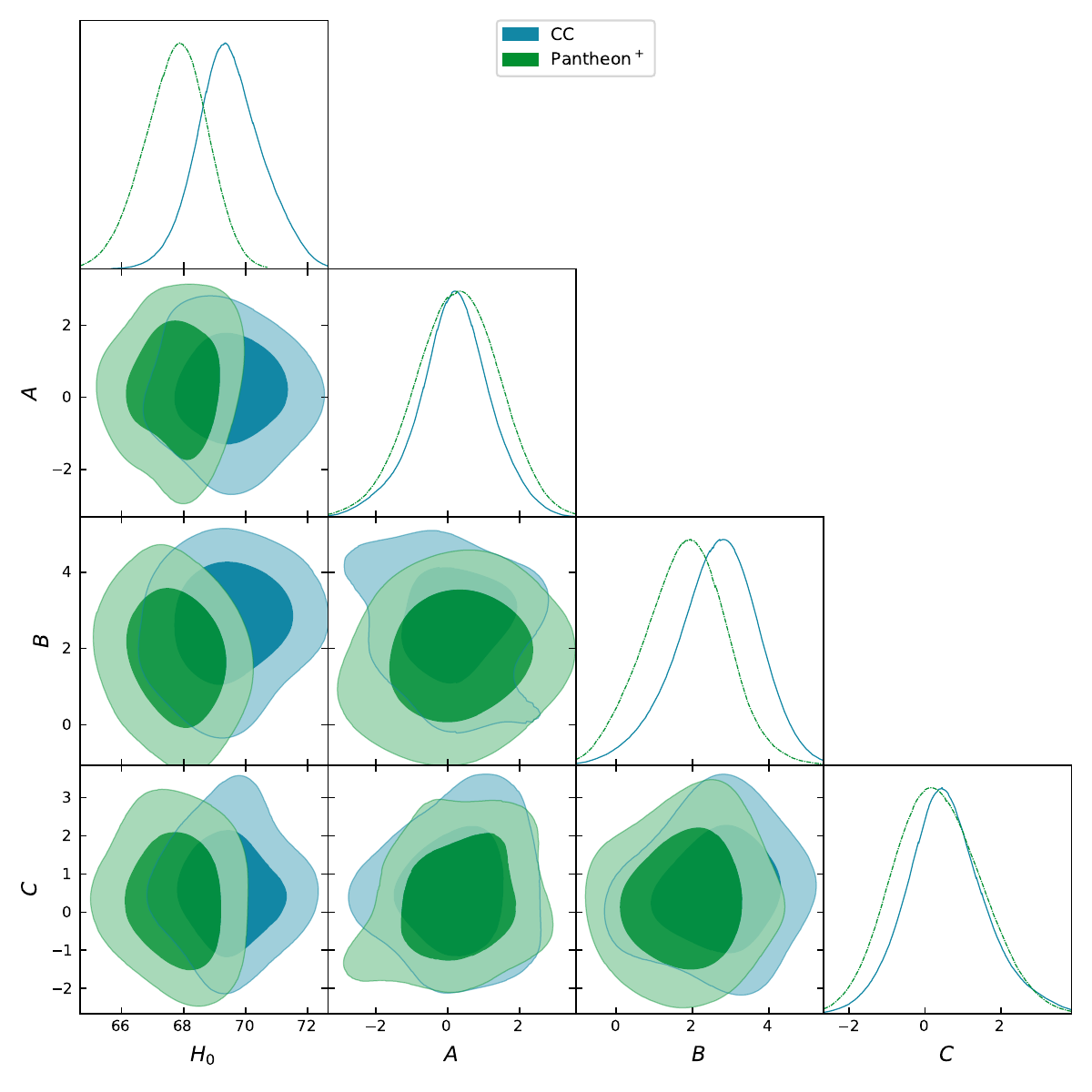}
        \caption{Contour plot obtained from CC and Pantheon$^+$ data set.}
        \label{HP}
    \end{subfigure}%
    \hfill
    \begin{subfigure}[b]{0.65\textwidth}
        \centering
        \includegraphics[width=105mm]{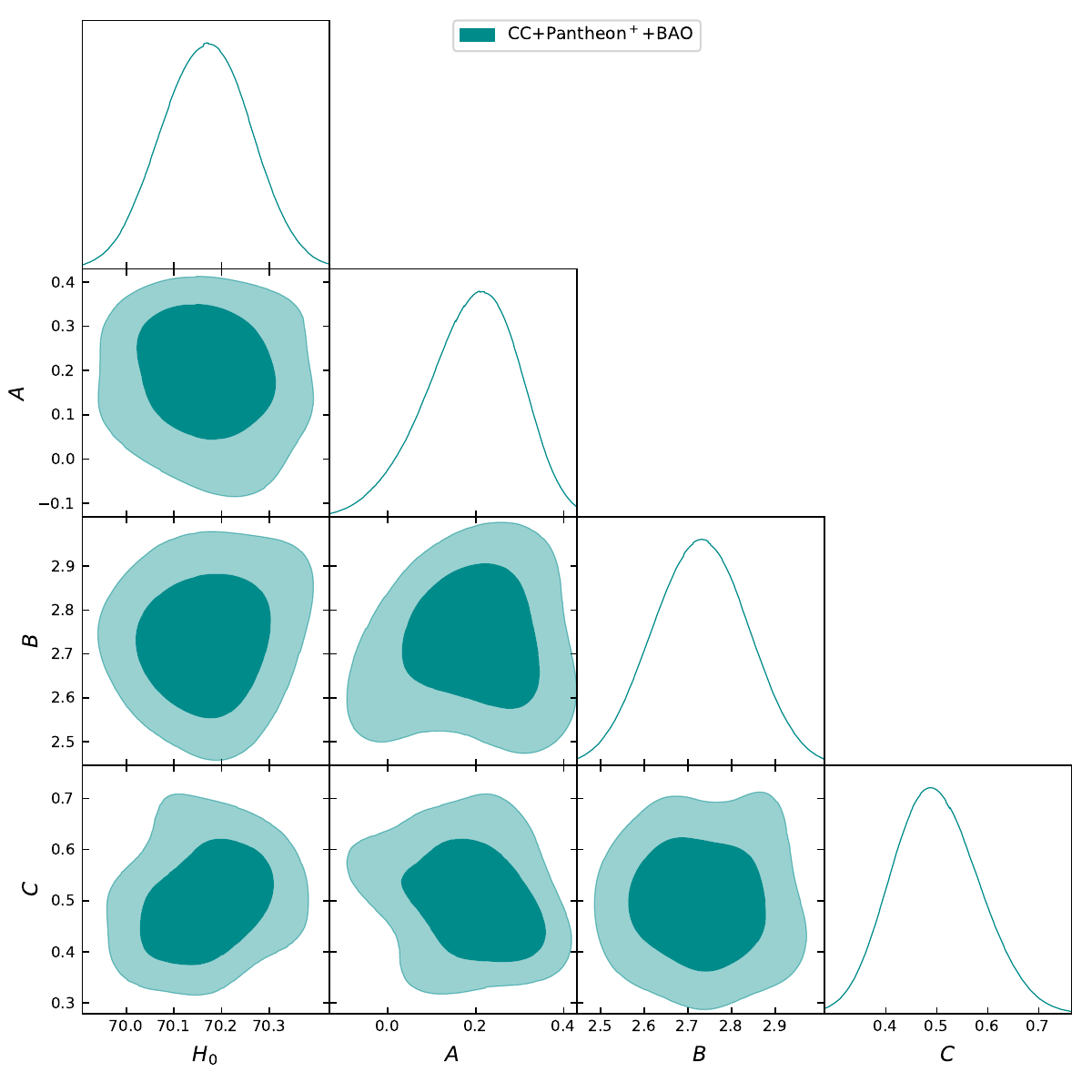}
        \caption{Contour plot obtained from $\text{CC}+\text{Pantheon}^{+}+\text{BAO}$ data set.}
        \label{HPB}
    \end{subfigure}%
    \caption{MCMC contour plot obtained from the observational data sets for 1$\sigma$ and 2$\sigma$ confidence interval.}
    \label{fig:I}
\end{figure*}

\noindent\textbullet{\textbf{BAO Data:}} Baryon Acoustic Oscillations (BAO) data have been gathered using SDSS, 6dFGS, and Wiggle Z surveys. The BAO data has been integrated with the following:
\begin{equation}\label{eq28}
d_{A}(z)=c \int_{0}^{z} \frac{d\tilde{z}}{H(\tilde{z})},
\end{equation}
\noindent where $d_{A}(z)$ is the comoving angular diameter distance, $c$ is the speed of light and $H(\tilde{z})$ represents the Hubble \\
\\\\
parameter. Also, the dilation scale can be obtained as, 
\begin{equation}\label{eq29}
D_{v}(z)=\left[ \frac{d_{A}^2 (z) c z }{H(z)} \right]^{1/3}.
\end{equation}
\noindent For the BAO data, the chi-square function ($\chi^{2}$) is employed, which can be expressed as
\begin{equation}\label{eq30}
\chi^{2}_{\text{BAO}} = X^{T} C_{\text{BAO}}^{-1} X,
\end{equation}
where $C_{\text{BAO}}^{-1}$ represents the inverse covariance matrix and the determination of $X$ is on the specific survey being analyzed \cite{Giostri2012}.

\noindent The free parameters $H_{0}$, $A$, $B$ and $C$ are constrained using the CC, Pantheon$^+$ and $\text{CC}+\text{Pantheon}^{+}+\text{BAO}$ data sets [Fig. \ref{fig:I}] and values are given in Table - \ref{Table:I}.
\begin{widetext}

 \begin{table}[H]
		\fontsize{10pt}{4pt}
		\addtolength{\tabcolsep}{0.5pt}
		\begin{center}
  \renewcommand{\arraystretch}{1.2}
			\begin{tabular}{ccccc}
				\hline
				\hline
				Data sets & $H_{0}$ & $A$ & $B$ & $C$ \\ [0.1cm] 
				\hline\hline\\[-0.1cm]  
				CC & $69.49\pm 1.10$ & $0.19\pm 0.99$ & $2.66^{+1.10}_{-0.97}$ & $0.55^{+0.95}_{-1.10}$\\[0.1cm] 
                Pantheon$^+$ & $69.5^{+2.10}_{-1.80}$ & $0.20^{+2.00}_{-2.20}$ & $2.70\pm 2.10$ & $0.5^{+2.20}_{-2.10}$\\[0.1cm]
                $\text{CC}+\text{Pantheon}^{+}+\text{BAO}$ & $70.16^{+1.00}_{-0.90}$ & $0.19^{+0.15}_{-0.13}$ & $2.73\pm 0.10$ & $0.49^{+0.08}_{-1.00}$\\[0.1cm]
				\hline
			\end{tabular}
		\caption{Best-fit values of parameter space using CC, Pantheon$^+$ and $\text{CC}+\text{Pantheon}^{+}+\text{BAO}$ data set.}\label{Table:I}
		\end{center}
	\end{table}
 \end{widetext}
The estimated value of the Hubble constant $H_{0}$ from our analysis, using the CC, Pantheon$^+$ and the combined $\text{CC} + \text{Pantheon}^{+} + \text{BAO}$ data sets are respectively $69.49$~km/s/Mpc, $69.50$~km/s/Mpc and $70.16$~km/s/Mpc. These results are consistent with current local measurements, $70.4\pm1.4$~km/s/Mpc \cite{Jarosik_2011_192_14}. This result is in mild tension with the value inferred from the Cosmic Microwave Background (CMB), which is approximately $67.4$~km/s/Mpc based on Planck data \cite{Aghanim_2020_641_A6}. The slight variations observed in our estimates reflect the impact of different datasets and highlight the ongoing challenges in resolving the Hubble tension.

In this section, we have reconstructed the cosmological model within the $f(Q)$ gravity framework and constrained the free parameters for Hubble parameter using MCMC analysis using different observational data sets. These constrained parameters and the $H(z)$ will be utilize in the next section to study the dynamical aspects of this reconstructed model by analyzing the time evolution of energy density and the equation of state, along with testing stability under perturbations. This analysis is essential for understanding the physical validity and long-term behavior of the model, ensuring that the reconstructed model align with observational data and remain stable over cosmic evolution.

\section{Model Analysis}\label{Sec:IV}
    \subsection{Dynamical behavior}
        Here, we will analyze the dynamical behavior of the Universe. Using the Eqs. \eqref{eq20}, \eqref{eq21}, \eqref{eq23} and Hubble parameter one can obtain the expression for the energy density and pressure for the model, which are expressed in \hyperref[Appendix]{Appendix}.
\begin{figure}[!htb]
    \centering
    \begin{subfigure}[b]{0.50\textwidth}
        \centering
        \includegraphics[width=83mm]{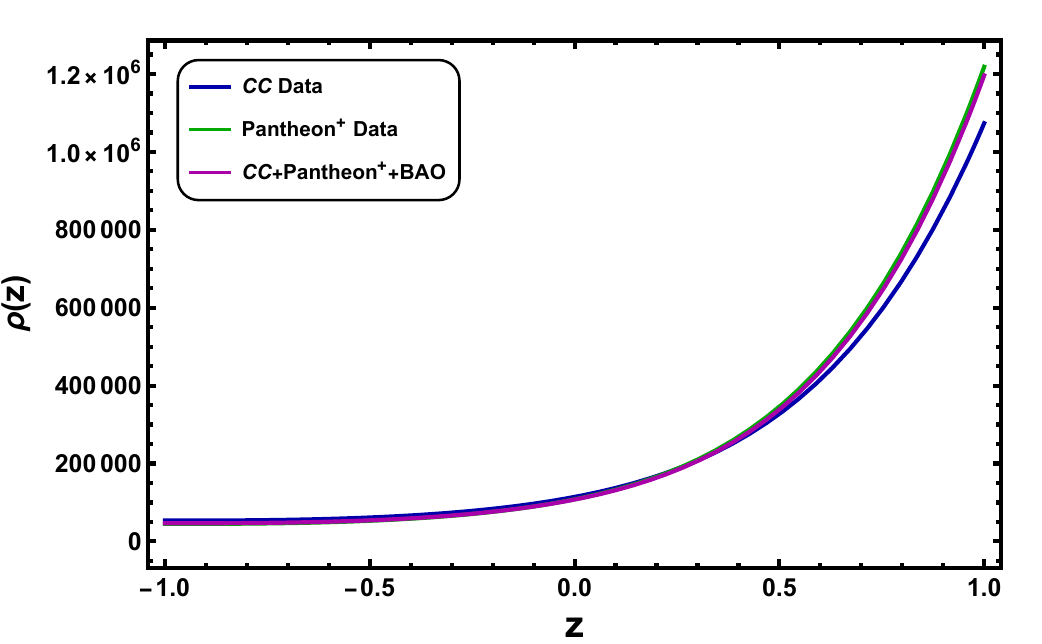}
        \caption{Evolution of energy density in redshift.}
        \label{rho1}
    \end{subfigure}%
    \hfill
    \begin{subfigure}[b]{0.50\textwidth}
        \centering
        \includegraphics[width=80mm]{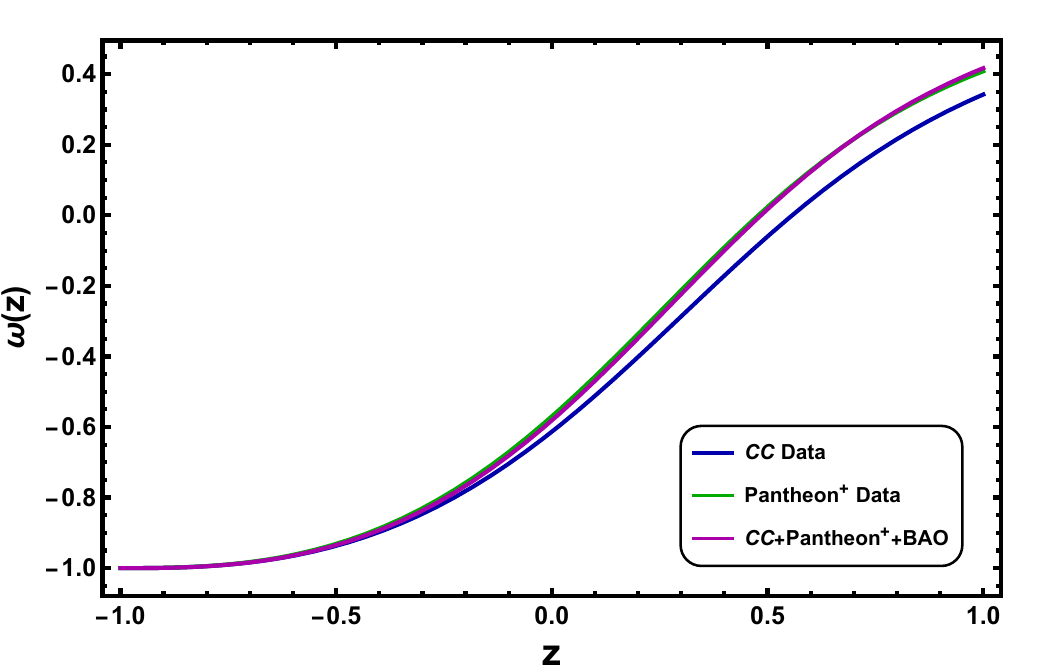}
        \caption{Evolution of EoS parameter in redshift.}
        \label{omega1}
    \end{subfigure}%
    \caption{The behavior of the dynamical parameters in redshift. The parameter scheme: $\alpha_{1}=1.20$, $\alpha_{2}= 0.50$, $\alpha_{3}= -0.0001$, $\beta_{1}=0.90$ and $\beta_{2}=0.01$.}
    \label{fig:II}
\end{figure}

Fig. \ref{rho1} shows that while the energy density decreases over time, it never completely vanishes. The current values of the EoS parameter are $\omega_{0} \simeq -0.63$, $\omega_{0} \simeq -0.59$, and $\omega_{0} \simeq -0.60$ for the constrained free parameters from the CC dataset, Pantheon$^{+}$ dataset, and the combined $\text{CC}+\text{Pantheon}^{+}+\text{BAO}$ dataset, respectively from Fig. \ref{omega1}. The energy density and EoS parameters are determined by the coefficients of the model, which influence their evolution. This model exhibits quintessence behavior at present.
\begin{figure}[!htb]
    \centering
    \begin{subfigure}[b]{0.50\textwidth}
        \centering
        \includegraphics[width=77mm]{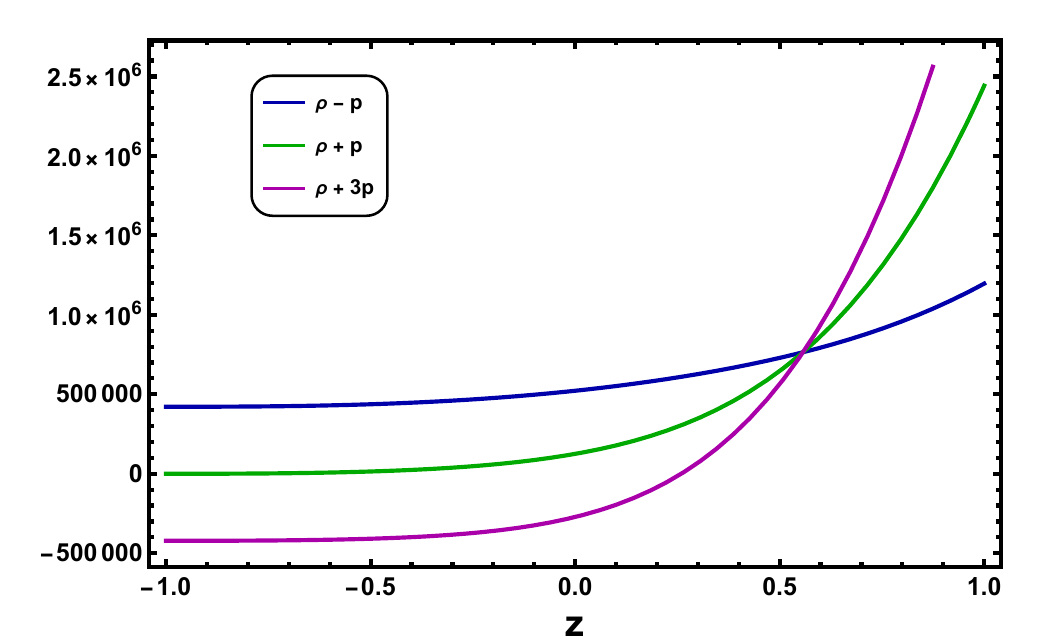}
        \caption{Energy conditions in redshift for CC data.}
    \end{subfigure}%
    \hfill
    \begin{subfigure}[b]{0.50\textwidth}
        \centering
        \includegraphics[width=80mm]{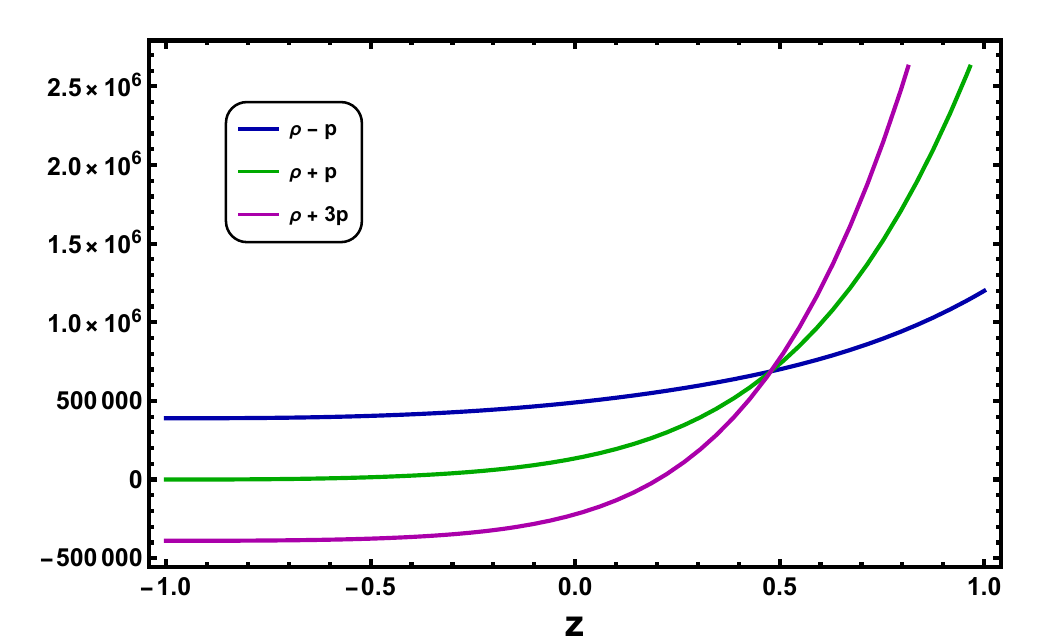}
        \caption{Energy conditions in redshift for Pantheon$^+$ data.}
    \end{subfigure}%
    \hfill
     \begin{subfigure}[b]{0.50\textwidth}
        \centering
        \includegraphics[width=80mm]{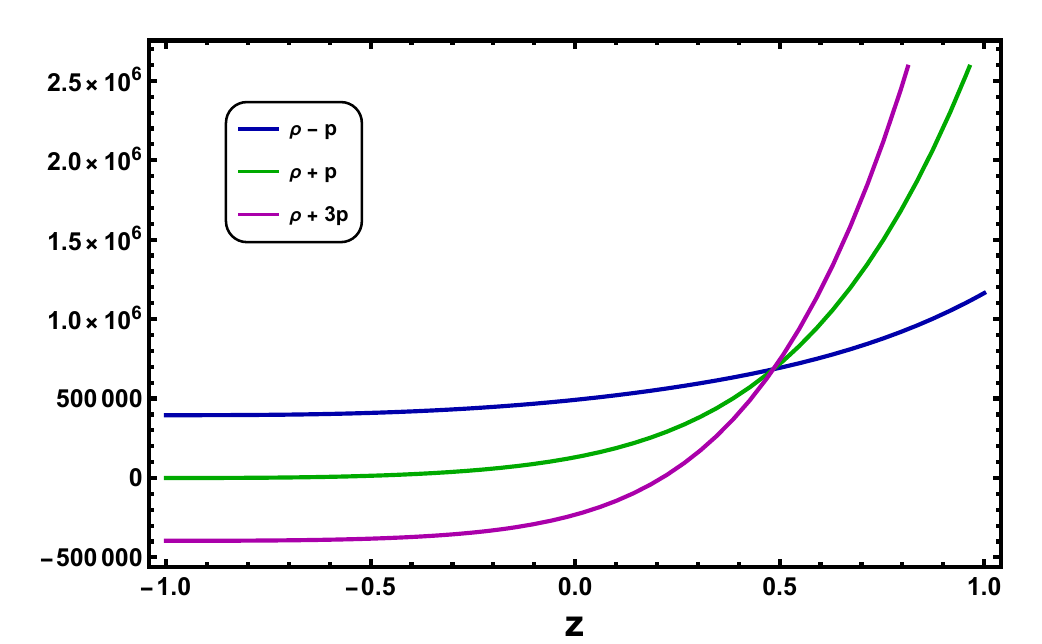}
        \caption{Energy conditions in redshift for $\text{CC}+\text{Pantheon}^{+}+\text{BAO}$ data.}
    \end{subfigure}%
    \caption{The behavior of energy conditions in redshift. The parameter scheme: $\alpha_{1}=1.20$, $\alpha_{2}= 0.50$, $\alpha_{3}= -0.0001$, $\beta_{1}=0.90$ and $\beta_{2}=0.01$.}
    \label{fig:III}
\end{figure}
In the study of cosmological models within modified gravity theories, examining energy conditions are crucial. Specifically, in the context of covariant $f(Q)$ gravity, we investigate the behavior of various energy conditions. The energy density must remain positive throughout cosmic evolution, and these energy conditions essentially serve as boundary conditions. The violation of the Strong Energy Condition due to dark energy suggests that these boundary conditions may not hold physical relevance. However, because of the fundamental casual structure of space-time, the gravitation attraction can be characterized by the energy conditions \cite{Capozziello2019}. Moreover, these boundary conditions play a significant role in shaping the cosmic evolution of the Universe \cite{Carroll2003}. As illustrated in Fig. \ref{fig:III}, across all datasets, the Null Energy Condition decreases from early to late time, remains positive throughout, but approaches zero in late time. In contrast, the Dominant Energy Condition remains positive at all times without any violation. The Strong Energy Condition is satisfied in the early Universe but is violated in the late epoch.

\subsection{Analyzing Stability with Scalar Perturbations} 
Our focus in this section will be on perturbations of homogeneous and isotropic FLRW metrics and their evolution, which ultimately determines whether cosmological solutions are stable in $f(Q)$ gravity. In order to study perturbations near the solutions $H(t)$ and $\rho(t)$, let us consider small deviations from the Hubble parameter and the evolution of the energy density defined as \cite{Dombriz2012, Farrugia2016},
\begin{equation}\label{eq31}
    H(t)\rightarrow H(t)(1+\delta), \quad \rho\rightarrow\rho(1+\delta_{m}),
\end{equation}
where $\delta$ and $\delta_{m}$ represent the isotropic deviation of the Hubble parameter and the matter over density, respectively. In this case, $H(t)$ and $\rho(t)$ represent zero order quantities [in some references, these are sometimes designated as $H_{0}(t)$ and $\rho_{0}(t)$, but this notation is avoided here in order to distinguish from quantities evaluated at present], which satisfy Eqs. \eqref{eq20}, \eqref{eq21} and continuity equation. The perturbation of the function $f$ and its derivatives are
\begin{equation*}
    \delta f = f_{Q}\delta Q,\quad \delta f_{Q} = f_{QQ}\delta Q, \quad \delta f_{QQ} = f_{QQQ}\delta Q.
\end{equation*}
where $\delta x$ represents the first-order perturbation of the variable $x$. Here $\delta Q = (-12H^{2}+9\gamma_{1}H)\delta$. In this way, the perturbed equations of Eq. \eqref{eq20} and continuity equation become,
\begin{eqnarray}
    c_{1}\dot{\delta}+c_{0}\delta &=& \rho\delta_{m},\label{eq32}\\
    \dot{\delta_{m}}+3H(1+\omega)\delta &=& 0.\label{eq33}
\end{eqnarray}
where
    \begin{align*}
    c_{0} =& ~6f_{Q}H^{2}-72f_{QQ}H^{4}+108f_{QQ}H^{3}-\frac{81f_{QQ}H^{2}}{2}\\
    &+36f_{QQ}H\dot{H}-\frac{27f_{QQ}\dot{H}}{2}-\frac{243}{2}f_{QQQ}H\dot{H}\\
    &-216f_{QQQ}H^{3}\dot{H}+324f_{QQQ}H^{2}\dot{H},\\
    c_{1} =& ~18f_{QQ}H^{2}-\frac{27f_{QQ}H}{2}.
\end{align*}
By solving Eqs. \eqref{eq32} and \eqref{eq33}, one can determine the stability of a particular FLRW cosmological solution in the context of $f(Q)$ gravity with connection-III. Since Eq. \eqref{eq32} has a linear character, the solution for $\delta(t)$ can generally be split into two branches: the first one corresponds to the solution of the homogeneous equation in \eqref{eq32}, which reflects perturbations induced by a particular gravitational Lagrangian. The second branch would correspond to the particular solution of that equation, which is merely affected by the growth of matter perturbations $\delta_{m}$. Considering the Eqs. \eqref{eq23}, \eqref{eq32}, \eqref{eq33} and the Hubble parameter, one can find the perturbation equations for the reconstructed model.

\begin{figure}[!htb]
    \centering
    \includegraphics[width=\linewidth]{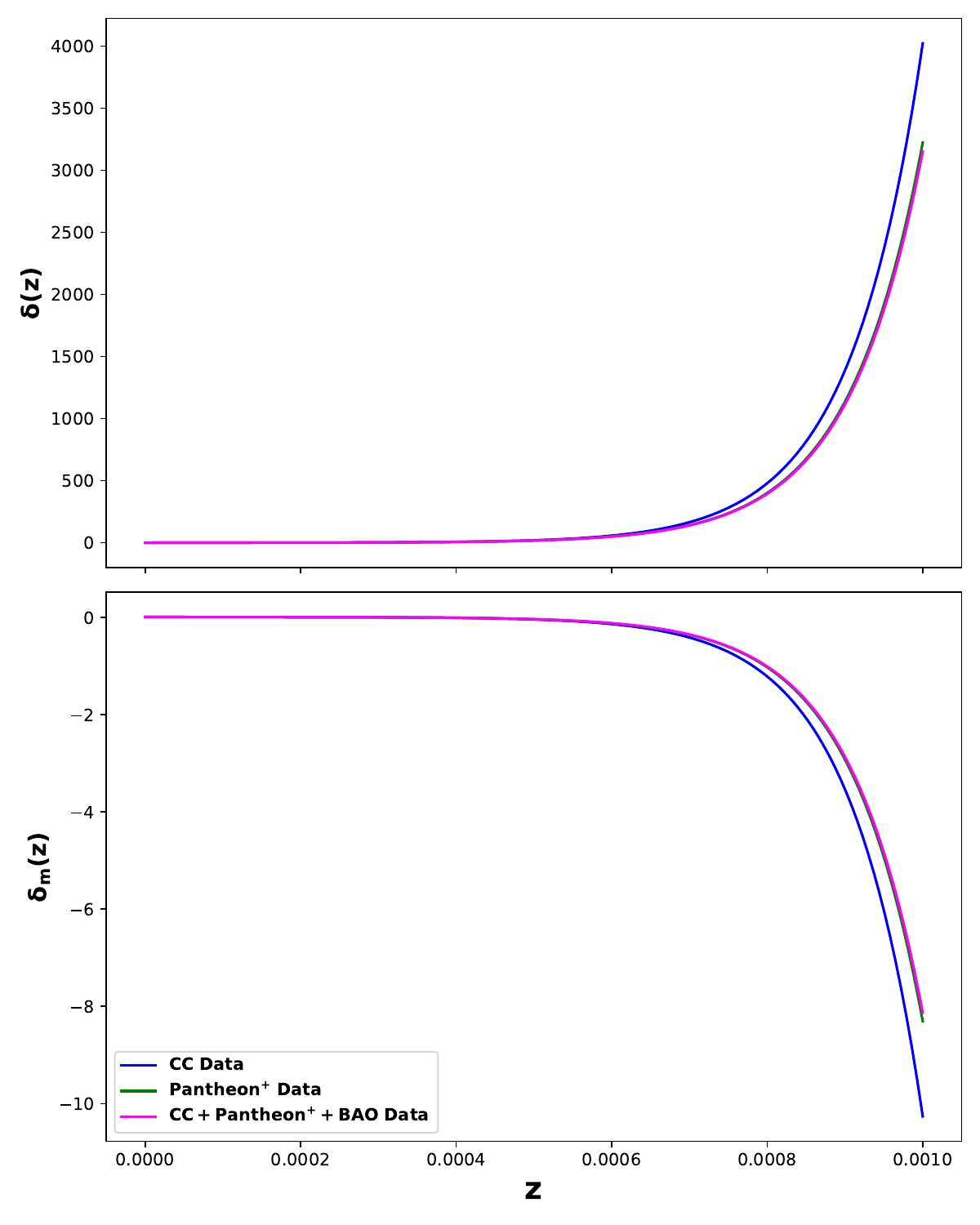}
    \caption{Evolution of $\delta$ and $\delta_{m}$ versus redshift. Herein, we set the initial conditions $\delta(0) = 0.1$ and $\delta_{m}(0)$ = 0.01.}
    \label{fig4}
\end{figure}
The analysis of cosmological perturbations $\delta$ and $\delta_{m}$ as illustrated in Fig. \ref{fig4} provides valuable insights into the behavior of these perturbations across different redshifts. Perturbation $\delta$ is compared against three data sets: CC dataset, Pantheon$^{+}$ dataset, and the combined $\text{CC}+\text{Pantheon}^{+}+\text{BAO}$ dataset. Similarly, perturbation $\delta_{m}$ is evaluated using the same data sets. Both perturbations exhibit small variations, demonstrating consistency across different observational constraints. Importantly, it can be seen that $\delta$ and $\delta_{m}$ converge to zero at the late epoch in the redshift, which confirms the stability of the model with respect to the Hubble parameter. This convergence highlights the robustness of the model in describing the expansion of the Universe over time.

\section{Conclusions}\label{Sec:V}
    In this work, we implemented the cosmological reconstruction method in $f(Q)$ gravity using non-trivial connections, specifically connection III in a symmetric teleparallel background. Connection III is characterized by a non-metricity scalar incorporating the Hubble parameter and an additional function, distinguishing it from other connection sets. This unique composition yields distinctive cosmological dynamics and potential stability concerns. The non-metricity introduced by connection III offers a novel geometric perspective beyond the curvature and torsion of GR, with implications for parallel transport and the evolution of the Universe. Through this approach, we derived an intriguing cosmological model, which is a noteworthy step in categorizing the association of cosmological models in modified gravity. Our model has been meticulously analyzed using the Hubble parameter $H(z)$ and various observational datasets, providing a robust representation of the expansion history of the Universe. The model exhibits quintessence behavior at the current epoch, characterized by a dynamic dark energy component that evolves over time. This behavior transitions smoothly to the $\Lambda$CDM model at late times, which is consistent with the observed accelerated expansion of the Universe. This dual behavior underscores the versatility and potential of $f(Q)$ gravity in addressing the cosmological constant problem. We have also thoroughly evaluated the energy conditions. The Null Energy Condition remains positive throughout the cosmic evolution, indicating that the energy density is non-negative and the speed of sound is real. The Dominant Energy Condition is consistently satisfied, ensuring that the energy density exceeds the pressure and energy propagation remains within causal bounds. The Strong Energy Condition, which is satisfied in the early Universe, is violated in the late epoch. This violation is consistent with the observed accelerated expansion and suggests a deviation from standard matter-dominated models.

    The stability analysis through scalar perturbations has confirmed the robustness of our model. The perturbations remain well-behaved, ensuring the model is free from instabilities that could otherwise undermine its physical viability. Our findings highlight the potential of $f(Q)$ gravity as a viable alternative to the $\Lambda$CDM model. This framework provides a comprehensive explanation for the current accelerating expansion of the Universe and offers new insights into the cosmological constant problem. The versatility of $f(Q)$ gravity in accommodating both quintessence and $\Lambda$CDM behaviors makes it a promising candidate for future cosmological studies. This study significantly adds to our comprehension of cosmological dynamics within the context of modified gravity theories. The outcomes achieved in this study create opportunities for additional research into the consequences of $f(Q)$ gravity and its ability to tackle important questions in modern cosmology.

\section*{Acknowledgement} 
    SVL would like to express gratitude for the financial support provided by the University Grants Commission (UGC) through the Senior Research Fellowship (UGC Reference No. 191620116597) to carry out the research work. BM acknowledges the support of the Council of Scientific and Industrial Research (CSIR) for the project grant (No. 03/1493/23/EMR II).

\begin{widetext}

\section*{Appendix}\label{Appendix}
\begin{align*}
    \rho =& ~\frac{-\sqrt{3}}{2(27\gamma_{1}^{2}-8Q)^{3/2}}\bigg(648  \beta_{1} \gamma_{1}^{2}  H^2-729  \beta_{1} \gamma_{1}^{4}-4374  \beta_{2} \gamma_{1}^{4} H^2-192  \beta_{1} H^2 Q+3240  \beta_{2} \gamma_{1}^{2}  H^2 Q+32  \beta_{2} Q^3-576  \beta_{2} H^2 Q^2\\[10pt] 
    & +576  \beta_{1} H \dot{H}+7776  \beta_{2} \gamma_{1}^{2}  H \dot{H}-1728  \beta_{2} H \dot{H} Q-32  \beta_{1} Q^2-108  \beta_{2} \gamma_{1}^{2}  Q^2-432  \beta_{1} \dot{H}-5832  \beta_{2} \gamma_{1}^{2}  \dot{H}+1296  \beta_{2} \dot{H} Q\\
    &+324  \beta_{1} \gamma_{1}^{2}  Q \bigg)-\frac{1}{2(27\gamma_{1}^{2}-8Q)}\bigg(96 \alpha_{3} H^2 Q^2 +48 \alpha_{2} H^2 Q -162 \alpha_{2} \gamma_{1}^{2}  H^2 -324 \alpha_{3} \gamma_{1}^{2}  H^2 Q +576 \alpha_{3} H \dot{H} Q-8 \alpha_{3} Q^3 \\[10pt]
    &+27 \alpha_{3} \gamma_{1}^{2}  Q^2 -1944 \alpha_{3} \gamma_{1}^{2}  H \dot{H}  +8 \alpha_{1} Q -432\alpha_{3} \dot{H} Q +1458 \alpha_{3} \gamma_{1}^{2}  \dot{H} -27 \alpha_{1} \gamma_{1}^{2}  \bigg)~,\\[10pt]
    p =& ~48 \alpha_{3} H^2 \dot{H}-3 \alpha_{2} H^2-6 \alpha_{3} H^2 Q-48 \alpha_{3} H \dot{H}-2 \alpha_{2} \dot{H}+9 \alpha_{3} \dot{H}-4 \alpha_{3} \dot{H} Q+\frac{\alpha_{3} Q^2}{2}-\frac{\alpha_{1}}{2}-\frac{1}{2} (\beta_{1}+\beta_{2}Q)\sqrt{81 \gamma_{1}^{2} -24 Q}\\[10pt]
    &+\frac{\sqrt{3}}{(27 \gamma_{1}^{2} -8 Q)^{3/2}}\bigg(384  \beta_{1} H \dot{H}-384  \beta_{1} H^2 \dot{H}+1152  \beta_{2} H^2 \dot{H} Q-5184  \beta_{2} \gamma_{1}^{2} H^2 \dot{H}+5184  \beta_{2} \gamma_{1}^{2} H \dot{H}-1152  \beta_{2} H \dot{H} Q\\[10pt]
    &+216  \beta_{2} \dot{H} Q-72  \beta_{1} \dot{H}-972  \beta_{2} \gamma_{1}^{2} \dot{H}\bigg)+\frac{1}{\sqrt{9 \gamma_{1}^{2} -\frac{8 Q}{3}}}\bigg(12 \beta_{1} H^2+36 \beta_{2} H^2 Q+8 \beta_{1} \dot{H}-81 \beta_{2} \gamma_{1}^{2}  H^2-54 \beta_{2} \gamma_{1}^{2}  \dot{H}\\[10pt]
    &+24 \beta_{2} \dot{H} Q-6 \beta_{2} Q^2-2 \beta_{1} Q+\frac{27 \beta_{2} \gamma_{1}^{2}  Q}{2}\bigg)~,
\end{align*}
where 
\begin{align}
    \dot{H} = -\frac{1}{2}ABH_{0}^2 (1+z)^B \left(1+\frac{A (1+z)^B}{\sqrt{A^2 (1+z)^{2 B}+C}}\right)~.
\end{align}
The Appendix presents detailed expressions for the variables $\rho$ and 
$p$, which are derived from a field equations describing the reconstructed model we are working with. These equations incorporate various parameters such as $\alpha_{1}$, $\alpha_{2}$, $\alpha_{3}$, $\beta_{1}$,$\beta_{1}$, and $\gamma_{1}$ along with dynamical quantities like the Hubble parameter 
$H$ and its derivative with respect to time $t$ (i.e. $\dot{H}$). Additionally, the equation for $\dot{H}$ is provided in the context of redshift $z$.

\end{widetext} 

\bibliographystyle{JCAP.bst}
\bibliography{Reference}
\end{document}